\ifdefined\pdfminorversion
\fi
\PassOptionsToPackage{table}{xcolor}
\documentclass[sigconf,authorversion]{acmart}

\usepackage{mathtools}
\usepackage{booktabs}
\usepackage{multirow}
\usepackage{pifont}
\usepackage{microtype}
\usepackage{siunitx}
\usepackage{xcolor}
\usepackage{colortbl}
\usepackage{enumitem}
\usepackage{xspace}
\usepackage{graphicx}
\usepackage{fontawesome5}
\usepackage{tabularx}
\usepackage{array}
\usepackage{float}
\usepackage{needspace}
\usepackage{xurl}
\usepackage[most]{tcolorbox}
\usepackage{hyperref}

\definecolor{autoblue}{HTML}{2563EB}
\definecolor{autobluelight}{HTML}{EEF5FF}
\definecolor{autoorange}{HTML}{EA6A19}
\definecolor{autoorangelight}{HTML}{FFF4EA}
\definecolor{autopurple}{HTML}{6D3CCB}
\definecolor{autopurplelight}{HTML}{F5F0FF}
\definecolor{tablehead}{HTML}{EAF2FF}
\definecolor{tablerow}{HTML}{F7FAFF}
\newcolumntype{Y}{>{\raggedright\arraybackslash}X}
\newcolumntype{Z}{>{\raggedleft\arraybackslash}X}
\newcolumntype{C}[1]{>{\centering\arraybackslash}p{#1}}
\newcolumntype{L}[1]{>{\raggedright\arraybackslash}p{#1}}
\newtcolorbox{promptbox}[2][]{enhanced,breakable,colback=autobluelight,colframe=autoblue,
  boxrule=0.7pt,arc=1.5mm,left=2mm,right=2mm,top=1.5mm,bottom=1.5mm,
  title={#2},fonttitle=\bfseries,coltitle=white,colbacktitle=autoblue,
  attach boxed title to top left={xshift=2mm,yshift=-2mm},
  boxed title style={arc=1mm,boxrule=0pt},before skip=8pt,after skip=8pt,#1}
\newtcolorbox{xrdcasebox}[3][]{enhanced,colback=white,colframe=autoblue!35,
  colbacktitle=autobluelight,coltitle=autoblue!55!black,boxrule=0.5pt,
  titlerule=0.4pt,arc=1.5mm,left=2mm,right=2mm,top=1.1mm,bottom=1.1mm,
  toptitle=0.7mm,bottomtitle=0.7mm,fonttitle=\small\bfseries,fontupper=\small,
  title={\strut #2\hfill{\normalfont\footnotesize\strut #3}},
  before upper={\setlength{\parskip}{3pt}},before skip=7pt,after skip=7pt,#1}
\tcbset{e2ecase/.style={colframe=autoorange!40,colbacktitle=autoorangelight,
  coltitle=autoorange!53!black}}
\newcommand{\casefield}[2]{\par\noindent\textbf{#1.}\enspace #2}

\copyrightyear{2027}
\acmYear{2027}
\setcopyright{cc}
\setcctype{by-nc-nd}
\acmConference[WSDM '27]{Proceedings of the Twentieth ACM International Conference on Web Search and Data Mining}{February 15--19, 2027}{Hong Kong, Hong Kong}
\acmBooktitle{Proceedings of the Twentieth ACM International Conference on Web Search and Data Mining (WSDM '27), February 15--19, 2027, Hong Kong, Hong Kong}
\acmDOI{10.1145/3862322.3863556}
\acmISBN{979-8-4007-3117-4/2027/02}
\newcommand{\method}{\textsc{AutoXRD}\xspace}
\newcommand{\bench}{\textsc{XRDBench}\xspace}
\newcommand{\benchqa}{\textsc{XRDBench-QA}\xspace}
\newcommand{\benchetwo}{\textsc{XRDBench-E2E}\xspace}
\newcommand{\rwp}{R_{\mathrm{wp}}}
\newcommand{\rexp}{R_{\mathrm{exp}}}

\newcommand{\ind}{\mathbb{I}}
\newcommand{\clip}{\operatorname{clip}}

\newcommand{\yes}{\textcolor{green!50!black}{\ding{51}}}
\newcommand{\no}{\textcolor{red!75!black}{\ding{55}}}
\title[AutoXRD: Autonomous LLM Agents and Evaluation for Powder XRD Analysis]{AutoXRD: Autonomous Large Language Model Agents and Comprehensive Evaluation for Powder X-Ray Diffraction Analysis}

\author{Yuetong Wu}
\authornote{Both authors contributed equally to this research.}
\orcid{0009-0009-8774-0573}
\affiliation{%
  \institution{The Chinese University of Hong Kong}
  \city{Hong Kong}
  \country{China}
}
\affiliation[obeypunctuation=true]{%
  \department{State Key Laboratory of Quantum Information Technologies and Materials},
  \institution{The Chinese University of Hong Kong},
  \city{Hong Kong},
  \country{China}
}
\email{yuetongwu@link.cuhk.edu.hk}
\author{Maojun Sun}
\authornotemark[1]
\authornote{Corresponding author.}
\orcid{0009-0006-4629-7489}
\affiliation{%
  \department{Department of Data Science and Artificial Intelligence}
  \institution{The Hong Kong Polytechnic University}
  \city{Hong Kong}
  \country{China}
}
\email{mj.sun@connect.polyu.hk}
\renewcommand{\shortauthors}{Wu and Sun}

\begin{abstract}
Powder X-ray diffraction (XRD) is a foundational technique for structural and phase characterization in materials science, yet automating its analysis and evaluating it end-to-end remains challenging. Doing so requires an agent to interpret diffraction patterns, run refinement software, manage strongly coupled parameters in a physically meaningful order, and assess both numerical fit quality and physical validity. We propose AutoXRD, an autonomous large language model (LLM) agent framework that organizes powder XRD analysis into refinement stages, uses diffraction diagnostics to select actions, and checks the resulting fits against predefined crystallographic and physical criteria. We further introduce XRDBench with two complementary tracks. XRDBench-QA contains 100 tasks assessing scientific reasoning and decision-making, whereas XRDBench-E2E contains 34 end-to-end workflows requiring data interpretation, crystallographic software execution, iterative refinement, evidence preservation, and reporting. Across 1,340 runs of ten recent LLMs, models average only 57.8 out of 100, falling from 61.9 on XRDBench-QA to 53.7 on XRDBench-E2E. At the task level, they perform best on refinement-history assessment and result acceptance, but remain substantially weaker on refinement-action selection, phase quantification, indexing, and Rietveld refinement. Among individual models, GPT-5.6 Sol achieves the highest overall score of 81.1, GPT-5.6 Terra scores highest on XRDBench-E2E (81.0), and GPT-5.6 Luna provides the best score-cost trade-off.
Moreover, ablation studies show that all six AutoXRD components consistently improve performance, validating our framework design. Finally, execution-trace analysis reveals recurring failures in coupled-parameter control, quantitative reasoning, evidence preservation, and workflow termination, motivating stronger scientific constraints, uncertainty quantification, and more efficient planning.

\par\smallskip
\begingroup\centering
  \faGithub\ 
  \href{https://stephen-smj.github.io/AutoXRD/}
  {\texttt{https://stephen-smj.github.io/AutoXRD/}}
\par\endgroup

\end{abstract}



\begin{document}
% \raggedbottom
% \afterpage{\flushbottom}
\maketitle

\section{Introduction}
\begin{figure*}[htbp]
    \centering
    \includegraphics[width=\textwidth]{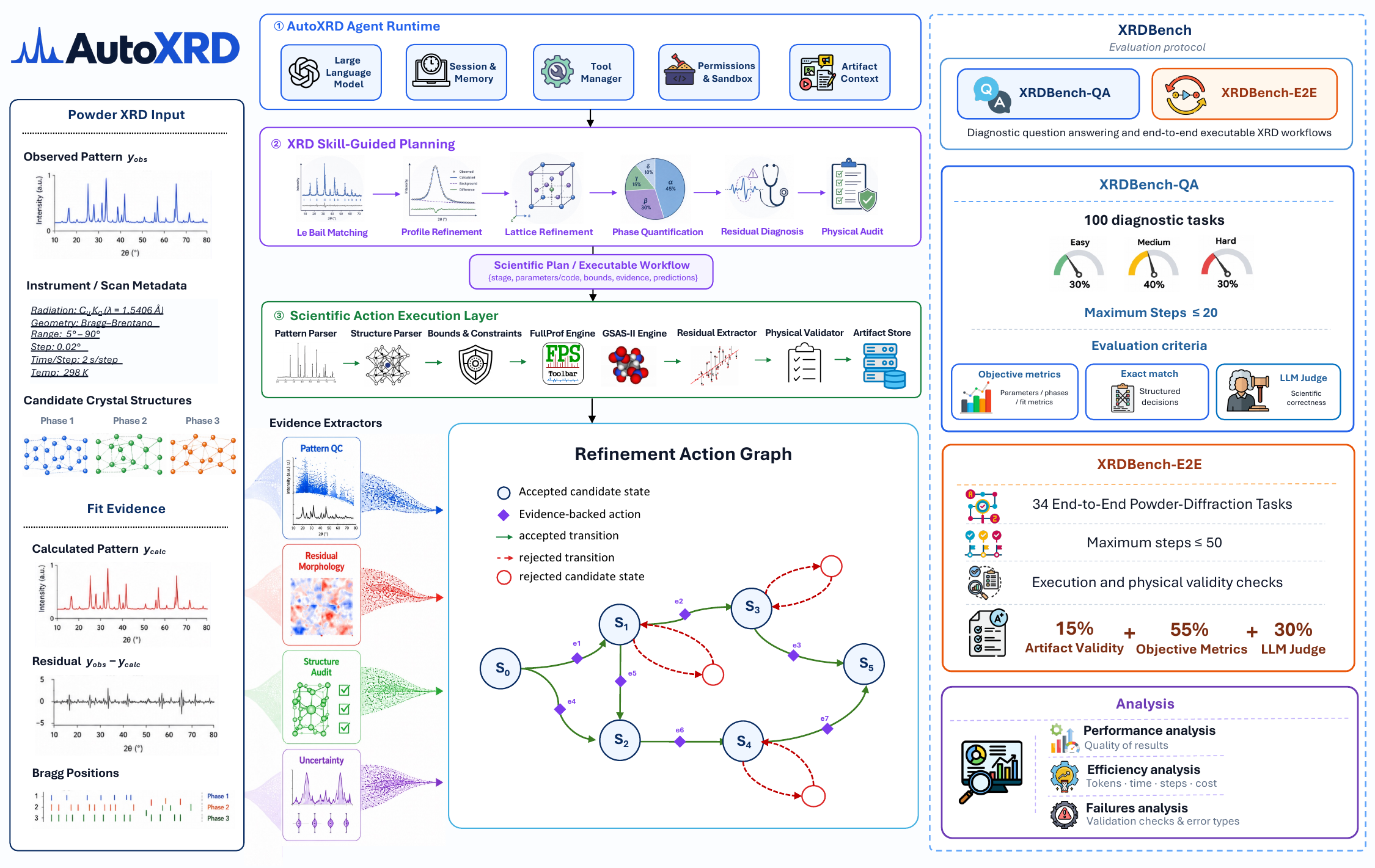}
    \caption{Overview of AutoXRD and the XRDBench evaluation protocol.
    AutoXRD combines an autonomous agent runtime, skill-guided powder-diffraction workflows, executable crystallographic backends, and evidence-based scientific validation. XRDBench evaluates diagnostic reasoning through XRDBench-QA and autonomous end-to-end XRD analysis through XRDBench-E2E.}
    \Description{System architecture diagram showing AutoXRD on the left, with an LLM runtime, ten XRD skills, executable crystallographic tools, and deterministic validation checks, and XRDBench on the right, with XRDBench-QA and XRDBench-E2E evaluation tracks.}
    \label{fig:autoxrd-overview}
\end{figure*}

{\setlength{\emergencystretch}{1em}%
Powder X-ray diffraction (XRD) projects three-dimensional reciprocal-space information onto a one-dimensional intensity profile.
Peak overlap, background, instrumental aberrations, preferred orientation, finite crystallite size, strain, and phase mixtures can therefore produce similar observations.
Rietveld refinement addresses overlap by fitting the complete profile \cite{rietveld1969,young1993}, but refinement cannot be treated as unconstrained curve fitting.
The parameters governing the crystal structure, specimen, and instrument are strongly coupled; successful practice depends on releasing them in a physically prescribed order and inspecting the observed, calculated, and difference patterns after each stage \cite{toby2006,toby2024}.\par}

These requirements expose two related gaps in current language-agent research.
First, while tool-using language agents can combine reasoning with external tool execution \cite{yao2023react,schick2023toolformer} and automate specific workflows in chemistry and experimentation \cite{bran2024chemcrow,boiko2023coscientist}, an unconstrained tool-use loop offers no guarantee of scientific validity.
For instance, a valid tool call may simultaneously free correlated parameters that cannot be uniquely determined, a lower fitting error ($\rwp$) may result from an incorrect structural or peak-shape assumption, and a fluent report may hide unphysical atomic occupancies or an unaccounted-for crystalline phase.
Second, existing evaluations of diffraction agents focus on specific applications, such as crystal structure prediction, single-crystal alignment, and Rietveld refinement. These studies demonstrate what agents can do within particular workflows, but offer limited guidance on how different models perform across powder-diffraction tasks such as indexing, phase identification and quantification, and refinement. Comparing models across these tasks calls for a benchmark built on measured X-ray and neutron powder patterns, with common requirements for software execution and output files and task-specific measures of scientific accuracy.

For iterative refinement, we formulate the workflow as a directed refinement action graph. Specifically, each node stores a complete fit state, whereas each edge encodes a structured action along with its supporting diagnostics and falsifiable predictions. When an action is proposed, numerical software computes the candidate next state, and deterministic checks, rather than the language model, decide whether to accept the transition. Consequently, the resulting refinement history retains both accepted and rejected transitions, thereby supporting failure diagnosis and policy distillation.

Figure~\ref{fig:autoxrd-overview} summarizes the proposed \method.
At runtime, the language model maintains session context and memory, plans with ten XRD-specific skills, and invokes permissioned tools within a sandbox.
The skills translate domain procedures into executable scientific plans that specify the refinement stage, parameters or code, numerical bounds, required evidence, and expected outcomes.
The scientific action layer parses diffraction patterns and structures, enforces parameter constraints, executes FullProf, GSAS-II, or task-appropriate numerical routines, and stores the resulting artifacts. The system evaluates pattern quality, residual structure, crystallographic validity, and parameter uncertainty to accept or reject candidate transitions in the refinement action graph.

The right panel of Figure~\ref{fig:autoxrd-overview} presents \bench, designed to address the evaluation gap through two complementary tracks.
XRDBench-QA contains 100 tasks that assess scientific reasoning and decision-making, whereas XRDBench-E2E contains 34 end-to-end workflows requiring data interpretation, crystallographic software execution, iterative refinement, evidence preservation, and reporting. Both tracks report performance, efficiency, and failure behavior under a common evaluation framework.
Using \bench, we evaluate ten recent LLMs in 1,340 agent runs.
Average performance decreases from 61.9 on \benchqa to 53.7 on \benchetwo, and the two track scores are only moderately correlated ($r=0.56$, $\rho=0.65$), showing that diagnostic competence does not reliably translate into autonomous execution.
Models perform best on refinement-history assessment and result acceptance, but remain substantially weaker on refinement-action selection, phase quantification, indexing, and Rietveld refinement.

To summarize, our main contributions are:
\begin{itemize}[leftmargin=*,itemsep=2pt,topsep=3pt]

    \item \textbf{An autonomous agent for powder diffraction analysis.}
    We introduce \method, which uses XRD-specific skills to plan analyses and executes them with FullProf, GSAS-II, or task-specific code. For structured refinement steps, \method provides rule-based checks that test whether fit diagnostics change as predicted and whether the resulting fit meets crystallographic and physical criteria. 
    % Its refinement log can record both accepted and rejected steps, along with the corresponding diagnostics and output files. 

    \item \textbf{A two-track benchmark for powder diffraction agents.}
    We introduce \bench, with 100 \benchqa tasks covering refinement decisions, interpretation of diffraction data, parameter recovery, and phase analysis. \benchetwo contains 34 workflows spanning eleven task families on measured X-ray and neutron powder diffraction data. The end-to-end evaluation checks whether agents run the analysis and produce the required files, and assesses the numerical accuracy and physical validity of their results.

    \item \textbf{A large-scale empirical study of LLMs for XRD.}
    We evaluate ten recent LLMs across 1,340 agent runs, 22,135 model turns, and 625.6M tokens.
    Our analysis reveals that diagnostic capability does not reliably translate into autonomous execution. Furthermore, failure analysis identifies refinement-action selection, phase quantification, indexing, and Rietveld refinement as key bottlenecks.

\end{itemize}

\section{Background and Related Work}
\begin{table*}[hbpt]
\centering
\caption{Comparison of language agents for X-ray and neutron diffraction.
\yes{}/\no{} mark reported support or evaluation versus absence or no report.
Modules cover phase/structure analysis, executable refinement, fit diagnosis, scientific validation, and reporting.
Skills counts named procedures. Tools counts documented interfaces and utilities, including the six native tools for \method (see Appendix~\ref{app:resources}, Table~\ref{tab:agent-tools}).
E2E requires agent-controlled execution on a specific material. Mtls. counts simulated (sim.) or measured (real) inputs (E2E only for \method).
Runs counts model--task executions. LLM counts exclude judges and meta-optimizers. -- denotes not reported or not applicable.}
\label{tab:related-comparison}
\setlength{\tabcolsep}{2.6pt}
\renewcommand{\arraystretch}{1.18}
\scriptsize
\resizebox{\textwidth}{!}{%
\begin{tabular}{@{}lccccccclcccccc@{}}
\toprule
& \multicolumn{5}{c}{\textbf{Supported XRD modules}}
& \multicolumn{2}{c}{\textbf{Agent resources}}
& \multicolumn{7}{c}{\textbf{Agent evaluation scope}} \\
\cmidrule(lr){2-6}\cmidrule(lr){7-8}\cmidrule(lr){9-15}
System & Phase/Struct. & Refine & Fit Diag. & Valid. & Report & Skills & Tools & Task & QA & E2E & Mtls. & Tasks & Runs & LLMs \\
\midrule
LLM-CSP Agent \cite{cao2025llmcspagent}
  & \yes & \no & \no & \yes & \no & 0 & 5 & PXRD structure pred. & \no & \yes & $\sim$6.4K sim. & $\sim$6.4K & $\sim$6.4K & 1 \\
AI X-ray Scientist \cite{chen2026xrayagent}
  & \no & \no & \no & \yes & \yes & 0 & 3 & SCXRD alignment & \no & \yes & 2 real & 1 & 22 & 3 \\
Rongzai Agent \cite{li2026rongzai}
  & \no & \yes & \yes & \yes & \yes & -- & 4 & NPD refinement & \no & \yes & 5 real & 5 & 20 & 4 \\
AgentBuild-Rietveld \cite{shin2026agentbuild}
  & \no & \yes & \yes & \yes & \yes & 0 & 6 & PXRD refinement & \no & \yes & 3 real & 8 & 34 & 1 \\
NeuDiff Agent \cite{xiao2026neudiff}
  & \yes & \yes & \yes & \yes & \yes & 0 & 4 & SCND structure det. & \no & \yes & 1 real & 1 & 10 & 2 \\
\midrule
\textbf{\method (ours)}
  & \yes & \yes & \yes & \yes & \yes & \textbf{10} & \textbf{15} & PXRD/NPD, 11 families & \yes & \yes & 34 real & \textbf{134} & \textbf{1,340} & \textbf{10} \\
\bottomrule
\end{tabular}%
}
\end{table*}

\subsection{Rietveld Refinement}
Let $x_i$ denote the diffraction coordinate (e.g., $2\theta$, time of flight, or $d$-spacing), $y_i$ the observed intensity, $\sigma_i$ its standard uncertainty, and $\hat y_i(\theta)$ the profile calculated from parameters $\theta$.
Classical Rietveld refinement minimizes weighted least squares \cite{rietveld1969}:
\begin{equation*}
  \theta^* = \arg\min_{\theta\in\Theta}
  \chi^2(\theta),\qquad
  \chi^2(\theta)=\sum_{i=1}^{n}w_i\bigl[y_i-\hat y_i(\theta)\bigr]^2,
\end{equation*}
where $w_i=\sigma_i^{-2}$ when uncertainties are available.
The vector $\theta$ includes scale, background, zero shift, lattice, profile, atomic coordinates, displacement factors, occupancies, preferred orientation, and microstructure terms.
Le Bail profile matching can align the cell, background, and peak profile before atomic parameters are exposed \cite{lebail1988}.
For a fully crystalline mixture with all relevant phases included in the refinement model, phase weight fractions can be derived from refined scale factors and phase-specific $ZMV$ terms, where $Z$, $M$, and $V$ denote formula units per cell, formula mass, and unit-cell volume \cite{hill1987qpa}.
Preferred orientation and line shape require specialized models \cite{dollase1986,tch1987}.

No scalar fit index establishes structural correctness.
For example, the weighted-profile factor
\begin{equation*}
 \rwp = 100\sqrt{\frac{\sum_i w_i(y_i-\hat y_i)^2}{\sum_i w_i y_i^2}}
\end{equation*}
depends on data quality and uncertainty conventions; a wrong model on weak data may achieve a lower value than an excellent model on precise data \cite{toby2006}.
This motivates explicit checks of residual structure, parameter correlations, chemistry, bounds, and unresolved peaks.

\subsection{Automation of XRD Analysis}
Efforts to automate crystallographic engines such as FullProf \cite{rodriguez1993fullprof} and GSAS-II \cite{toby2013gsasii} have progressed from high-throughput scripting (SrRietveld) \cite{tian2013srrietveld} and rule-based expert systems (AutoFP) \cite{cui2015autofp} to reinforcement learning (PowderBot) \cite{feng2019powderbot}, black-box optimization (BBO-Rietveld) \cite{ozaki2020bbo}, derivative-guided recipe selection \cite{toby2024}, learned parameter prediction \cite{mun2026mlrietveld}, and robotic laboratory integration \cite{yotsumoto2024are}.
In parallel, automated phase identification, supported by expanding structural and synthetic repositories such as COD \cite{grazulis2012cod}, opXRD \cite{hollarek2026opxrd}, SimXRD-4M \cite{cao2025simxrd}, and SIMPOD \cite{rincon2025simpod}, has evolved from supervised deep classifiers \cite{lee2020deepxrd,lee2021augmentation} and probabilistic autonomous workflows (XCA) \cite{maffettone2021xca} to physics-informed frameworks that couple candidate search with refinement verification, including CrystalShift \cite{chang2025crystalshift}, Dara \cite{fei2026dara}, and RADAR-PD \cite{taniai2026radar}.
Most recently, end-to-end pipelines such as AGAPI-XRD \cite{campbell2026agapi} and MatDiffract \cite{wang2026matdiffract} have chained retrieval or structure generation directly with full-pattern refinement.
However, these algorithmic pipelines rely on fixed heuristics or black-box search; when confronted with ill-conditioned initial states or subtle fit anomalies, they lack the flexible reasoning required to dynamically diagnose residual signatures, regulate parameter turn-on sequences, and coordinate multi-stage workflows.

\subsection{Tool-Using and Scientific Agents}
ReAct interleaves language reasoning with actions \cite{yao2023react}, while Toolformer learns when and how to call APIs \cite{schick2023toolformer}.
AgentBench \cite{liu2024agentbench}, $\tau$-bench \cite{yao2025taubench}, MLAgentBench \cite{huang2024mlagentbench}, DSAEval \cite{sun2026dsaeval}, and RStatBench \cite{sun2026dare} show that endpoint success alone can hide long-horizon failures and inconsistency.
Scientific-agent work includes chemistry agents \cite{bran2024chemcrow}, autonomous laboratory orchestration \cite{boiko2023coscientist}, and broad research automation \cite{lu2024aiscientist,sun2026survey,sun2026lambda,sun2026rejoinder,sun2026rejoinder2}.
ScienceAgentBench argues for authentic, task-level assessment before claims of end-to-end discovery \cite{chen2025scienceagentbench}.

Recent large language model (LLM) agents attempt to address this flexibility gap by enabling dynamic scientific tool use, yet existing crystallographic agents remain constrained in both operational scope and evaluation rigor (Table~\ref{tab:related-comparison}).
Domain-specific agents typically target isolated workflow segments: the multimodal LLM-CSP agent \cite{cao2025llmcspagent} predicts crystal symmetry and structures without executing iterative profile refinement, whereas the AI X-ray Scientist \cite{chen2026xrayagent} focuses on hardware-level beamline alignment.
Closer to powder diffraction analysis, Rongzai \cite{li2026rongzai}, AgentBuild-Rietveld \cite{shin2026agentbuild}, and NeuDiff \cite{xiao2026neudiff} interface with GSAS-II or neutron reduction suites to automate structure refinement and fit diagnosis.
Despite demonstrating feasibility, these studies test only their native architectures on small, case-specific suites, such as a single repeated workflow in NeuDiff, five neutron tasks in Rongzai, and eight X-ray cases in AgentBuild.
Conversely, while OPENXRD \cite{vosoughi2026openxrd} scales evaluation to 217 crystallographic questions across 74 foundation models, it operates purely as a static question-answering benchmark without testing executable software invocation, iterative state control, or physical validity.

% \method extends task-level evaluation to powder crystallography, where physical and mathematical constraints govern the validity of refinement results.
% It differs from generic agents by making action schemas, falsifiable predictions, stage order, coupled-parameter exclusions, mandatory physical checks, and rejected steps part of the executable protocol.

By contrast, \method supports a broader range of powder diffraction analyses within a single framework, including phase analysis, indexing, and Rietveld refinement. It combines XRD-specific procedures with crystallographic software and task-specific code, and provides rule-based checks for structured refinement steps. To compare models across these workflows, \bench provides end-to-end tasks spanning eleven families and using measured X-ray and neutron powder patterns. A separate QA track assesses individual XRD decisions and analyses.

% \cyan{Consequently, the field lacks both a principled control framework to prevent unconstrained agent divergence during software execution and a standardized benchmark to disentangle reasoning failures from execution breakdowns.
% Our work bridges this divide by introducing a diagnostics-governed control architecture alongside \bench, which unifies 100 fine-grained diagnostic tasks and 34 executable end-to-end workflows across 11 crystallographic families to systematically evaluate 10 diverse LLM agents under a single protocol.}

\section{Problem Formulation}
\subsection{Refinement States and Structured Actions}
We formalize autonomous XRD analysis as a sequential decision-making process over structured crystallographic states. Specifically, each problem instance is defined as
\begin{equation*}
  \mathcal{I}=(\mathcal{D},\mathcal{M},\mathcal{C}),\quad
  \mathcal{D}=\{(x_i,y_i,\sigma_i)\}_{i=1}^{n},
\end{equation*}
where $\mathcal{M}$ is instrument metadata and $\mathcal{C}$ is a set of candidate structures.
At step $t$, the agent observes
\begin{equation*}
 s_t=(\theta_t,q_t,r_t,v_t,c_t,\mathcal{O}_t),
\end{equation*}
where $q_t=(R_{\mathrm{wp},t},R_{\mathrm{exp},t},\mathrm{GoF}_t)$ contains fit scalars, $r_t$ contains residual features, $v_t$ is a set of physical violations, $c_t$ describes complexity and correlations, and $\mathcal{O}_t$ indexes backend output files.
Here $R_{\mathrm{wp}}$ is the weighted-profile residual, $R_{\mathrm{exp}}$ is the statistically expected residual, and $\mathrm{GoF}=R_{\mathrm{wp}}/R_{\mathrm{exp}}$ denotes goodness of fit.

AutoXRD supports two complementary execution modes.
In the validated refinement path, the agent proposes a structured action
\begin{equation*}
 a_t=(k_t,g_t,P_t,B_t,\mathcal{Z}_t,\Pi_t),
\end{equation*}
where $k_t$ is an action kind, $g_t$ a refinement stage, $P_t$ a parameter subset, $B_t$ optional bounds, $\mathcal{Z}_t$ cited diagnostics, and $\Pi_t$ falsifiable predictions.
Each supporting diagnostic is formulated as $(f, \nu, \tau, \mathrm{src})$, comprising the feature name, observed value, decision threshold, and data source.
A prediction $\pi=(m,d,\delta)$ specifies metric $m$, required direction $d\in\{-1,+1\}$, and minimum effect $\delta\ge0$.
In the native code-agent path, actions may instead read, create, edit, or execute task files.
These broader actions are retained verbatim in the trajectory and are evaluated through validation checks for execution, artifacts, and physical validity.

For a structured refinement action, the numerical backend computes a candidate state and its associated output files:
\begin{equation*}
  (s_{t+1}',\mathcal{O}_{t+1})=F(s_t,a_t;\mathcal{D}).
\end{equation*}
The protocol then checks whether the objective improves, whether the fit diagnostics change as predicted, and whether the candidate state meets the physical and statistical criteria.

\subsection{Refinement Action Graph}
A refinement action graph is $G=(V,\mathcal{E}^+,\mathcal{E}^-)$.
Each $s\in V$ is a fit state, $\mathcal{E}^+$ contains accepted actions, and $\mathcal{E}^-$ contains rejected actions.
Let $H(s,a,s')\in\{0,1\}$ denote hard admissibility and
\begin{equation*}
 M(s,a,s')=\prod_{\pi\in\Pi(a)}
 \ind\!\left[d_\pi\bigl(m_\pi(s')-m_\pi(s)\bigr)\ge\delta_\pi\right]
 \label{eq:mechanism}
\end{equation*}
denote support for the action's expected outcome.
The candidate child is accepted only if
\begin{equation*}
 \mathrm{Accept}(s,a,s')=H(s,a,s')\,M(s,a,s')
 \ind\!\left[U(s')-U(s)>0\right]=1.
 \label{eq:accept}
\end{equation*}
Acceptance requires both checks to pass and the utility to improve.

\section{AutoXRD}
\subsection{Agent Architecture}
Figure~\ref{fig:autoxrd-overview} shows how \method plans and carries out a powder diffraction analysis. The controller manages the session, tool access, and resource budget. Using the supplied data and results from earlier steps, the LLM draws on ten XRD skills to plan its next action. These skills guide pattern and structure checks, refinement setup and staging, residual analysis, and final review.

The execution layer carries out the plan by reading the required files and running GSAS-II, FullProf, or task-specific code. For structured refinement steps, \method provides rule-based checks on parameter choices before execution and on diagnostic changes and physical validity afterward. The LLM uses the returned results and any check findings to decide whether to adjust the plan, continue the analysis, or report the result. The run history retains the attempted steps, their outputs, and the reasons for accepting or rejecting them, so the final report can explain how the result was obtained.

In this architecture, the LLM selects hypotheses and plans workflows, established crystallographic engines perform nonlinear least-squares optimization, and deterministic verification routines enforce domain invariants.

\subsection{Deterministic Residual Diagnostics}
For residuals $e_i=y_i-\hat y_i$, the feature extractor estimates robust scale
\begin{equation*}
 \hat\sigma_e=\frac{\operatorname{median}_i|e_i-\operatorname{median}(e)|}{0.67448975}
\end{equation*}
and standardized residuals $z_i=(e_i-\operatorname{median}(e))/\hat\sigma_e$.
Let $\tilde e_i=e_i-\bar e$, where $\bar e=n^{-1}\sum_i e_i$.
The extractor reports normalized absolute residual, lag-one autocorrelation
\begin{equation*}
 \rho_1=\frac{\sum_{i=1}^{n-1}\tilde e_i\tilde e_{i+1}}
 {\sum_{i=1}^{n}\tilde e_i^2},
\end{equation*}
low/high-angle signed bias, the fraction of points in contiguous regions with $|z_i|\ge5$, and an unexplained-peak ratio.
These features make diagnoses testable.
For example, once specimen displacement and other calibration effects have been ruled out, a nearly constant offset in peak positions can motivate zero-shift refinement. Before refining the zero shift, the agent specifies how much the peak offset should decrease. The observed change in peak positions tests this prediction, while $\rwp$ measures agreement across the full diffraction pattern.
A lower $\rwp$ indicates a better overall fit, while support for the zero-shift explanation requires the peak-position bias to decrease by at least the predicted amount.

\subsection{FullProf PCR Compilation}

For structured refinement with FullProf, \method generates a PCR input file from an existing template. The compiler uses the template comments to locate the numerical codewords that determine which parameters vary and which share a refinement variable. It checks that the template and the requested action are supported before assigning new codewords.

Let $P$ denote the parameters selected by an action of kind $k$, and let $\Gamma(k)$ denote the parameters allowed for that action. The compiler expands $P$ to include all members of any existing codeword group containing a selected parameter. The expanded set, $\operatorname{cl}(P)$, is subject to the condition
\begin{equation*}
  \operatorname{cl}(P) \subseteq \Gamma(k).
\end{equation*}
The new PCR file preserves the constraints within each selected group and holds the remaining cataloged parameters fixed.

\subsection{Refinement Checks}

For a structured refinement step, the action check verifies that the proposed adjustment is allowed at the current stage and that actions such as occupancy refinement specify parameter bounds. It rules out refining occupancy and isotropic displacement factors together, as well as refining the zero shift jointly with lattice parameters or wavelength.

After execution, the candidate state $s'_{t+1}$ is compared with the current state $s_t$. The admissibility check $H$ requires a valid action and a fit that introduces no new physical violations. It allows at most a $2\%$ relative increase in $\rwp$ and a $0.02$ increase in the unexplained-peak ratio. When parameter correlations are available, the maximum absolute correlation is limited to $0.98$. The prediction check $M$ separately tests whether the fit diagnostics change as predicted.

We compare the overall quality of the two states using
\begin{align*}
U(s)=-\biggl[&\frac{\rwp(s)}{\max(\rexp(s),\epsilon)}
+2R(s)+3Q(s)+0.002K(s) \\
&+10[C_{\max}(s)-0.90]_+ +10|v(s)|\biggr],
\end{align*}
where $R(s)$ is the residual score, $Q(s)$ is the unexplained-peak ratio, $K(s)$ is the parameter count, and $|v(s)|$ is the number of physical violations. The correlation term uses the maximum absolute parameter correlation $C_{\max}(s)$ and is set to zero when correlations are unavailable. Here $[x]_+=\max(x,0)$ and $\epsilon=10^{-9}$.

The step is accepted when $H=M=1$ and $U(s'_{t+1})>U(s_t)$.

\subsection{Complete Refinement History}
Each run record contains $(a_t,s_t,s_{t+1}',\mathrm{dec}_t,\mathcal{O}_{t+1})$, where $\mathrm{dec}_t$ is the acceptance decision and $\mathcal{O}_{t+1}$ lists the corresponding code, backend inputs, outputs, and saved diagnostics.
The history records the proposed action, supporting observations, before/after states, validation results, and output-file paths for every accepted or rejected step.
Retaining rejected attempts makes the final result reproducible and allows scientists to review which hypotheses were tried, why they failed, and how the accepted refinement path was obtained.

\begin{figure*}[t]
\centering
\includegraphics[width=\textwidth]{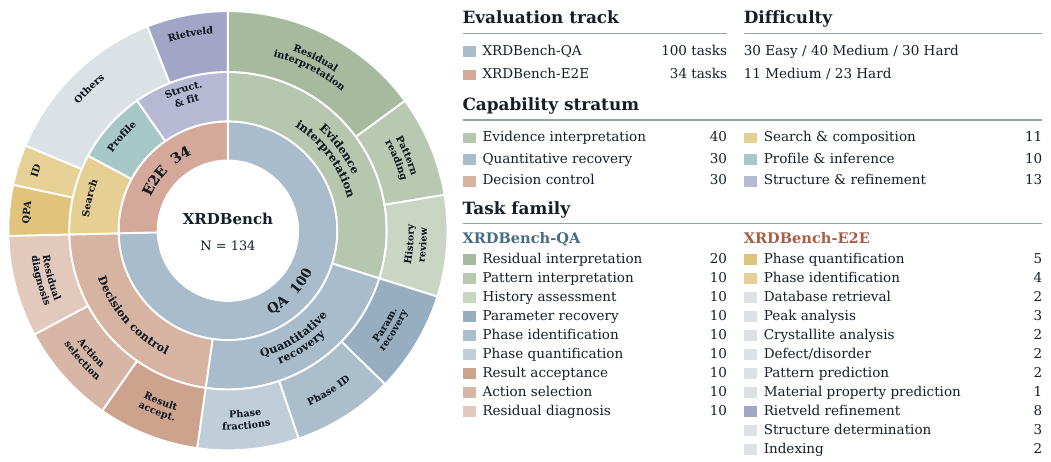}
\caption{Composition of XRDBench. The inner ring separates the 100 diagnostic tasks from the 34 executable end-to-end workflows; the middle ring groups related capabilities; and the outer ring shows nine XRDBench-QA capability categories and eleven XRDBench-E2E task families.}
\Description{A three-ring sunburst summarizes XRDBench by evaluation track, capability stratum, and task family, with separate difficulty counts.}
\label{fig:benchmark-composition}
\end{figure*}

% ===== Inlined from sections/xrdbench.tex =====
\section{XRDBench}
\subsection{Evaluation Tracks}

\bench evaluates agents on two separate sets of tasks. In the question-answering track, \benchqa, each case provides a question, the relevant diffraction data or refinement records, and a response format. Of the 100 cases, 30 are scored by exact matching of the selected options, 40 require structured responses assessed against reference answers and rubrics, and 30 require quantitative results and a report, both of which contribute to the score.

The end-to-end track, \benchetwo, contains 34 workflows based on measured diffraction data. The agent reads the given files, plans and runs an analysis, and submits the required output files and a report. For runs that pass the execution and physical checks, the score combines the quality of the output files, task-specific numerical metrics, and an assessment of the scientific report.

Figure~\ref{fig:benchmark-composition} shows how the tasks are distributed across tracks, capability groups, and task families.

\subsection{XRDBench-QA}
XRDBench-QA evaluates an agent's ability to make scientifically grounded decisions in XRD analysis.
Its 100 tasks cover nine capabilities spanning refinement decisions, diffraction data interpretation, residual diagnosis, and quantitative recovery.
Seventy tasks use deterministically generated inputs and ground truth, while the remaining 30 are constructed from experimental quantitative phase analysis (QPA) data reported by the IUCr and phase-identification cases adapted from Dara \cite{madsen2001qarr,fei2026dara}.
Each public task provides the relevant diffraction pattern or refinement history together with a required response format, whereas ground-truth answers, numerical targets, and scoring rubrics are retained by the evaluator.

For an objective metric $m$, baseline $b$, and oracle $o$, higher- and lower-is-better utilities are
\begin{equation*}
 u_{\uparrow}(m)=\clip\!\left(\frac{m-b}{o-b},0,1\right),\quad
 u_{\downarrow}(m)=\clip\!\left(\frac{b-m}{b-o},0,1\right).
\end{equation*}
Parameter recovery combines mechanism detection and normalized errors; QPA combines phase-set F1, fraction error/closure, amorphous accuracy, and report quality; phase identification combines phase-set F1, indexed coverage, unknown handling, and report quality.
With normalized case scores $u_j^{(Q)}\in[0,1]$, the QA track score is $S_Q=\sum_{j=1}^{100}u_j^{(Q)}$.

\subsection{XRDBench-E2E}
XRDBench-E2E evaluates agents on 34 powder diffraction workflows spanning 11 task families, including phase quantification, indexing, structure determination, and Rietveld refinement. The workflows use 30 Cu K$\alpha$ patterns from two opXRD sources \cite{hollarek2026opxrd} and four neutron diffraction examples distributed with FullProf. Of the 34 workflows, 11 are classified as Medium and 23 as Hard.

Each task provides diffraction data and instrument information in an isolated workspace without network access. The workspace includes XRD procedures, GSAS-II, FullProf, and numerical libraries. The agent examines the supplied files, selects and runs a numerical procedure, and revises the analysis as needed based on intermediate results. It submits a report together with the executed code and the required analysis outputs.

The execution check uses the run records and output files to verify that the agent examined the supplied inputs and ran the reported analysis. A separate physical check assesses whether the results satisfy the crystallographic constraints of the task. For task $j$, we set $g_j=1$ if both checks pass and $g_j=0$ otherwise. The score combines three components in $[0,1]$: output file validity ($\phi_{A,j}$), task-specific numerical accuracy ($\phi_{M,j}$), and the LLM judge's scientific assessment of the report and supporting results ($\phi_{J,j}$). The numerical component assesses, for example, estimated phase fractions in phase quantification and the recovered unit cell in indexing. We compute the task score and average it across all 34 workflows:
\begin{equation*}
\begin{aligned}
S_{E,j} &= 100g_j\left(0.15\phi_{A,j}+0.55\phi_{M,j}+0.30\phi_{J,j}\right),\\
S_E &= \frac{1}{34}\sum_{j=1}^{34} S_{E,j}.
\end{aligned}
\end{equation*}

\subsection{Score Aggregation and Run Records}
We report $S_Q$ and $S_E$ separately and give the two tracks equal weight in the overall score:
\begin{equation*}
S_{\mathrm{overall}} = \frac{S_Q+S_E}{2}.
\end{equation*}
For each run, we retain the execution records and output files and record wall-clock time, model turns, tool calls, and token use.

\begin{figure*}[htbp]
\centering
\includegraphics[width=\textwidth]{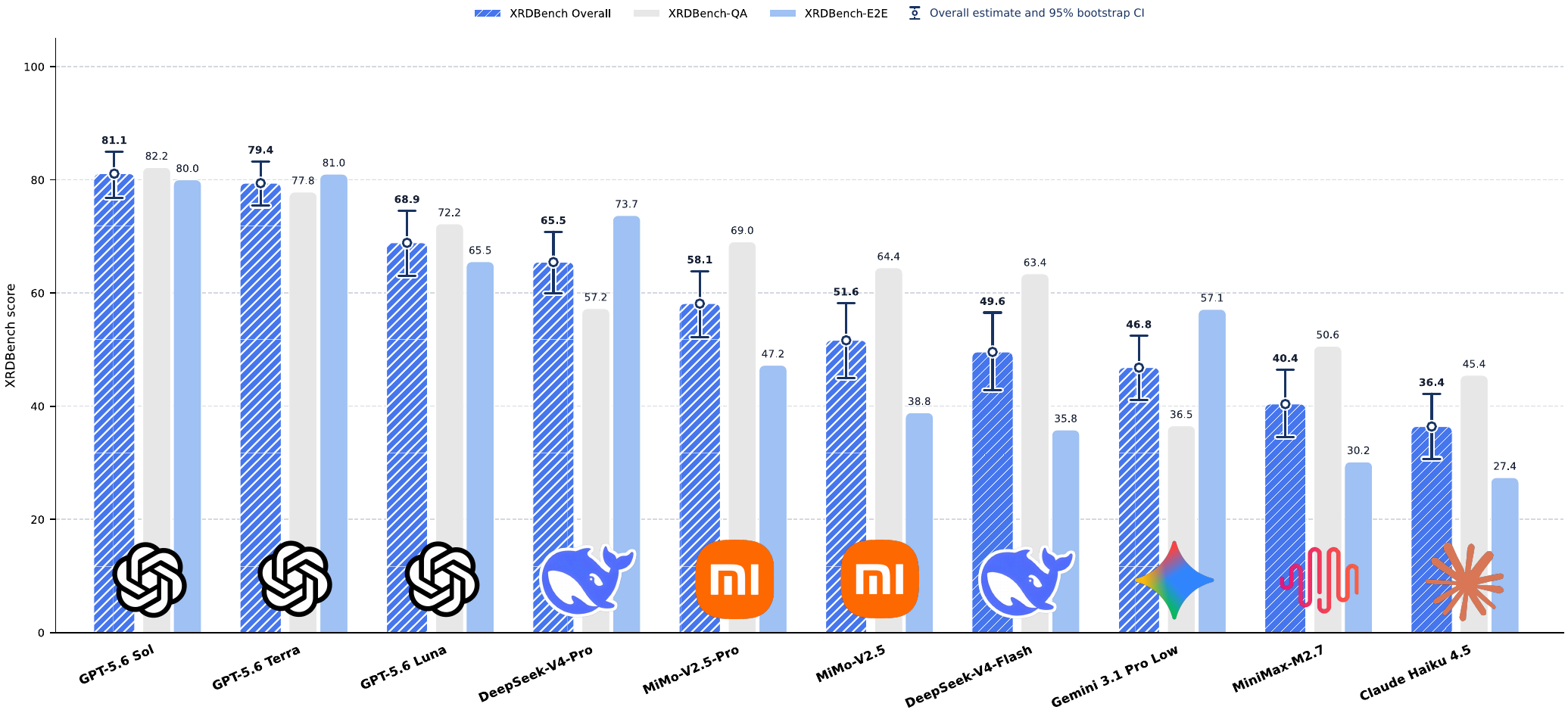}
\caption{Overall performance on XRDBench, ordered by the overall score.}
\Description{A grouped bar chart compares overall, QA, and end-to-end scores for ten models.}
\label{fig:overall-performance}
\end{figure*}

\section{Experiments}
\subsection{Experimental Setup}
We evaluate ten LLMs under the AutoXRD framework: GPT-5.6 Sol, GPT-5.6 Terra, GPT-5.6 Luna, MiMo-V2.5-Pro, MiMo-V2.5, DeepSeek-V4-Pro, DeepSeek-V4-Flash, MiniMax-M2.7, Claude Haiku 4.5, and Gemini 3.1 Pro Low.
Each \benchqa task allows at most 20 tool calls and 8,192 output tokens per turn, while each \benchetwo workflow allows at most 50 agent steps and 32,768 output tokens per turn.
For \benchetwo, GPT-5.6 Sol assigns the scientific-judgment score based on the submitted report and supporting results.
We use 20,000 task-level bootstrap replicates to estimate 95\% confidence intervals for track scores, overall scores, and pairwise model differences. In each replicate, we sample tasks with replacement separately within \benchqa and \benchetwo. For pairwise comparisons, we use the same resampled tasks for both models within each track. Additional details on agent tools, task-family mappings, prompts, and per-model results are provided in Appendices~\ref{app:resources}--\ref{app:results}.

\subsection{Research Questions}
We analyze 1,340 runs across ten models, totaling 22,135 model turns. We address five research questions: \textbf{RQ1: Performance on XRDBench-QA and XRDBench-E2E.} How do models perform on the two tracks, and which tasks remain most challenging? \textbf{RQ2: Cross-Track Model Rankings.} How closely do model rankings agree across the two tracks? \textbf{RQ3: Efficiency Analysis.} What are the trade-offs between model performance and runtime, model turns, token use, and inference cost? \textbf{RQ4: Failure Analysis.} What failures recur in end-to-end workflows? \textbf{RQ5: Component Ablation.} How does removing individual AutoXRD components affect diagnostic performance on \benchqa?

\begin{figure*}[htbp]
\centering
\includegraphics[width=\textwidth]{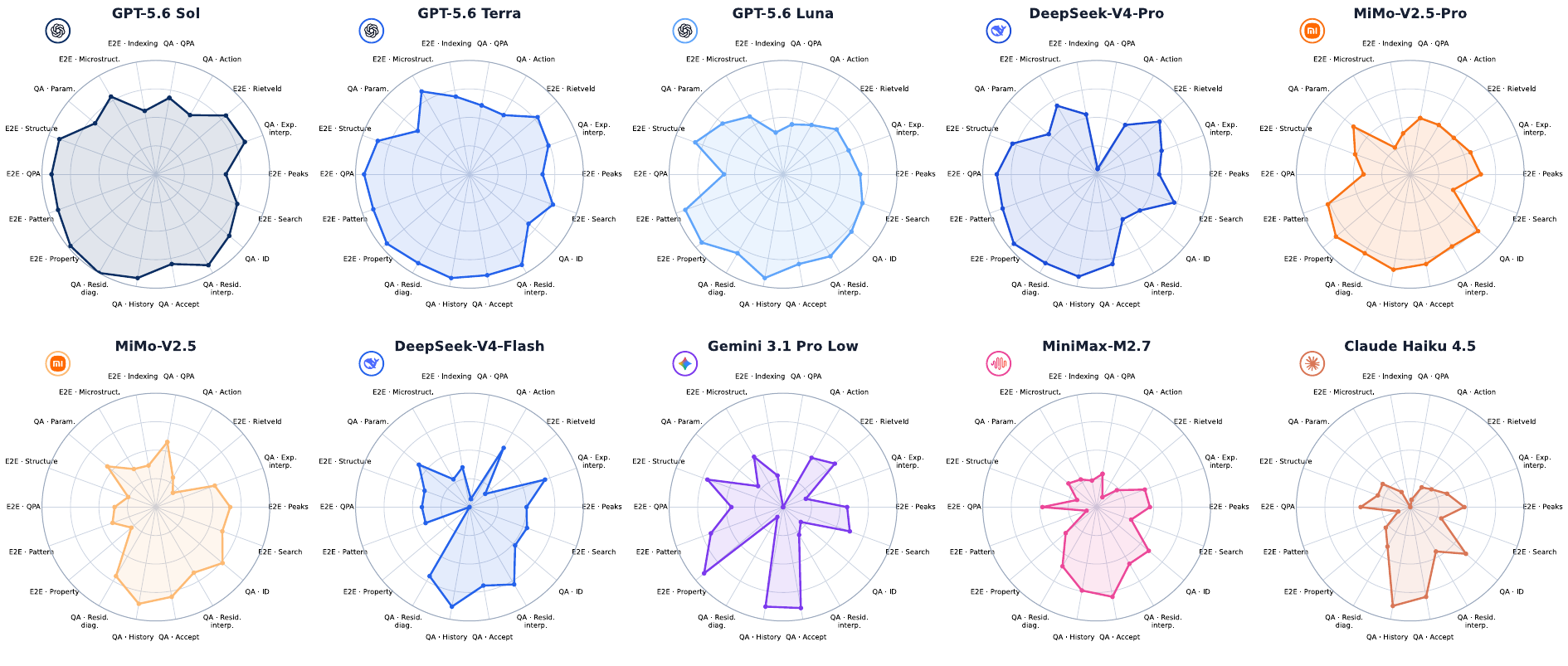}
\caption{Fine-grained performance across 18 dimensions: nine XRDBench-QA capabilities and nine end-to-end reporting domains. Each model has a separate 0--100 radar.}
\Description{Ten radar charts compare nine XRD capabilities and nine end-to-end reporting domains.}
\label{fig:combined-radar}
\end{figure*}

\begin{figure}[htbp]
\centering
\includegraphics[width=0.93\columnwidth]{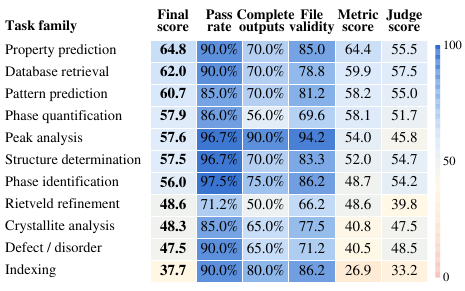}
\caption{XRDBench-E2E scores, validation pass rates, and deliverable completion rates for 11 task families, averaged over ten models.}
\Description{A heatmap compares final score, validation check pass rate, deliverable completion, artifact validity, objective metrics, and scientific judging across eleven end-to-end powder-diffraction task families.}
\label{fig:e2e-bottlenecks}
\end{figure}

\section{Results}

\begin{figure*}[t]
\centering
\includegraphics[width=\textwidth]{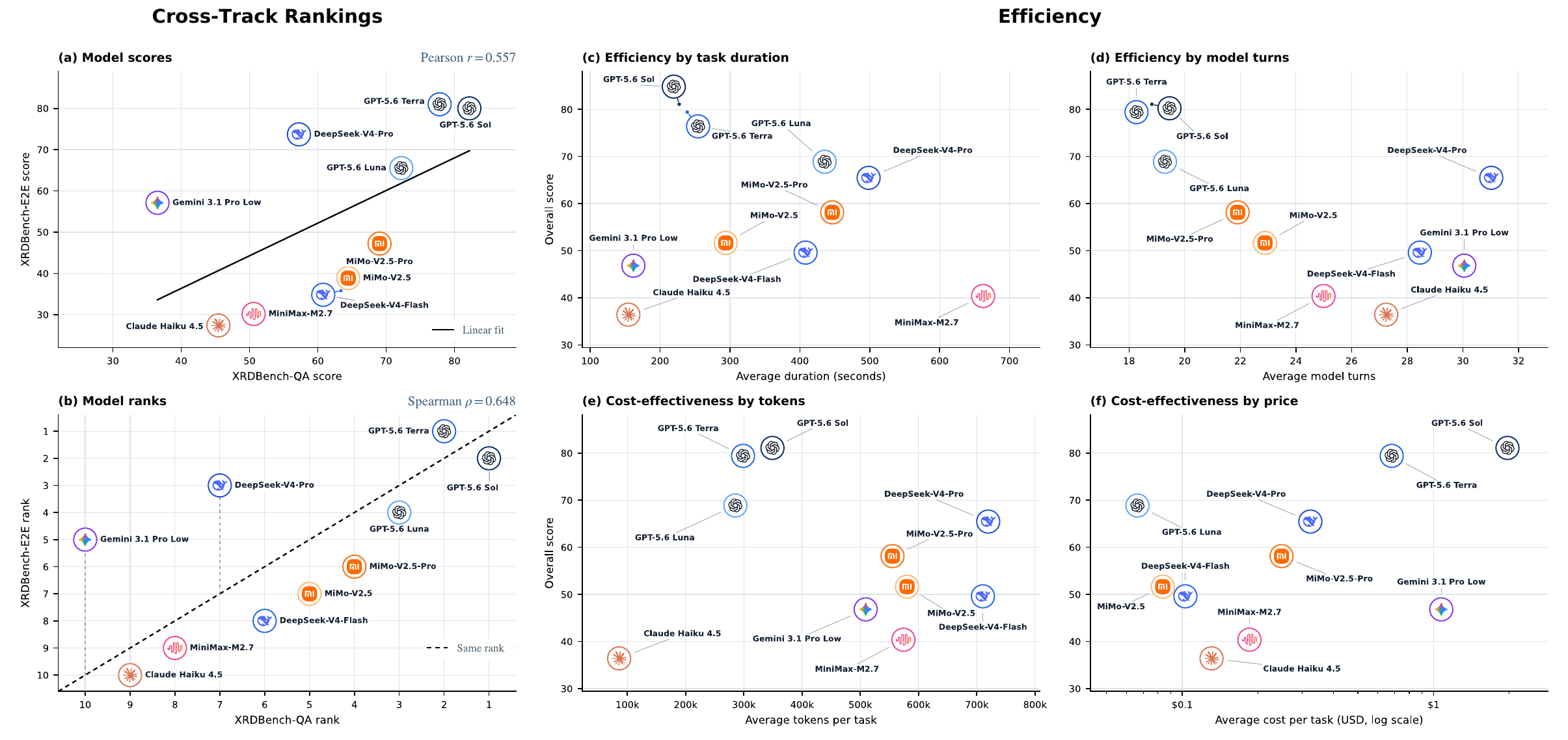}
\caption{Cross-track rankings and efficiency. (a,b) Scores and ranks across the two tracks. (c--f) Overall score versus mean runtime, model turns, tokens, and estimated inference cost. Resource means pool all 134 tasks per model.}
\Description{Six scatter plots compare ten models. The score panel shows a linear fit, and the rank panel shows an equal-rank diagonal. The other panels compare overall score with runtime, model turns, token use, and estimated inference cost. The cost axis is logarithmic.}
\label{fig:ranking-efficiency}
\end{figure*}

\subsection{RQ1: Performance on XRDBench-QA and XRDBench-E2E}
Autonomous end-to-end execution is substantially harder than standalone diagnostic reasoning (Figure~\ref{fig:overall-performance}).
Across the ten models, mean performance drops by 8.22 points from 61.89 on \benchqa to 53.67 on \benchetwo, with seven models scoring lower end-to-end.
GPT-5.6 Sol leads overall (81.09), followed closely by GPT-5.6 Terra (79.41) and GPT-5.6 Luna (68.87), leaving substantial room for improvement even among frontier models.

On \benchqa, capability-level results show a sharp gap between reviewing past refinements and making active scientific decisions (Figure~\ref{fig:combined-radar}).
Models reliably evaluate completed fits, averaging 88.0 on refinement-history assessment (nine of ten models $\ge 80$) and 81.0 on result acceptance.
Yet even this evaluation is easily misled when a scalar fit metric conflicts with residual diagnostics: in one case, nine of ten models accept a lower-$\rwp$ fit despite a new unexplained peak in the residual, and residual-diagnosis scores range widely from 10.0 (Gemini 3.1 Pro Low) to 100.0 (GPT-5.6 Sol).
Performance drops sharply on active control and quantitative tasks: action admissibility averages 44.0 (no model exceeds 60), all ten models fail on high-risk occupancy or joint zero-shift/lattice refinement cases, and quantitative phase analysis scores lowest overall (33.0; GPT-5.6 Sol leads at 68.2).

On \benchetwo, task-family means show a clear gap between workflow execution and scientific accuracy (Figure~\ref{fig:e2e-bottlenecks}).
Models score highest on prediction and retrieval workflows, including property prediction (64.79 on one workflow), database retrieval (62.03), and pattern prediction (60.70), but struggle on core structure- and parameter-recovery tasks.
Indexing is weakest by nearly ten points (mean 37.72), with seven models below 50 and GPT-5.6 Terra leading at 69.3 across its two workflows (Figure~\ref{fig:combined-radar}).
Crucially, indexing rarely fails during execution: 90\% of runs pass validation, 80\% produce complete outputs, and file validity reaches 86.2, yet numerical accuracy (26.9) and scientific-judgment scores (33.2) remain poor due to inaccurate unit cells and reflection assignments.
Peak analysis shows a similar gap, averaging 57.60 despite a 96.7\% validation pass rate and 90.0\% output completion.

Rietveld refinement, crystallite analysis, and defect/disorder analysis form a second low-scoring group (means 47.52--48.58).
Within this group, Rietveld refinement stands out for its low execution reliability, recording the lowest validation pass rate (71.25\%) and only 50\% complete outputs.
Among runs that pass validation, the contrast becomes clear: Rietveld refinement rises to 68.2, whereas indexing reaches only 41.9.
Thus, Rietveld refinement is heavily penalized by validation and completion failures, while indexing remains limited by solution accuracy even when workflows run cleanly.

\subsection{RQ2: Cross-Track Model Rankings}
Diagnostic strength on \benchqa only moderately predicts end-to-end execution on \benchetwo (Figure~\ref{fig:ranking-efficiency}(a,b)).
Across the ten models, track scores show a Pearson correlation of $r=0.557$ and a Spearman rank correlation of $\rho=0.648$, with pairwise orderings reversing for 10 of 45 model pairs.
These shifts go strongly in both directions: DeepSeek-V4-Pro and Gemini 3.1 Pro Low gain 16.46 and 20.58 points from \benchqa to \benchetwo (rising from seventh to third and tenth to fifth), whereas MiMo-V2.5-Pro, MiMo-V2.5, and DeepSeek-V4-Flash drop 21.8--27.6 points and fall two ranks each.

The two leaders also switch places: GPT-5.6 Sol tops \benchqa while GPT-5.6 Terra tops \benchetwo, leaving GPT-5.6 Sol ahead overall by a statistically non-significant 1.68 points (81.09 versus 79.41; paired 95\% CI $[-1.46,4.80]$).
Together, these rank reversals confirm that the two tracks measure complementary aspects of XRD capability: diagnostic reasoning in \benchqa versus multi-step tool execution and physically valid delivery in \benchetwo.
Reporting both tracks separately is therefore essential.

\subsection{RQ3: Efficiency Analysis}
Spending more runtime, tokens, or turns does not lead to higher accuracy on \bench (Figure~\ref{fig:ranking-efficiency}(c--e)).
Overall score has near-zero Spearman correlation with mean runtime ($0.08$) and token use ($-0.09$), and correlates negatively with model turns ($-0.64$), because weaker runs spend extra turns recovering from errors.
DeepSeek-V4-Pro, for instance, averages 31.0 turns and 719.7K tokens per task (1.7$\times$ and 2.4$\times$ GPT-5.6 Terra) yet scores much lower (65.47 versus 79.41).
Conversely, low resource use alone is not enough: Claude Haiku 4.5 is the fastest (155\,s/task) and uses the fewest tokens (85.9K tokens/task), yet ranks last overall (36.41).

Among the top three models, efficiency separates cleanly by cost and speed (Figure~\ref{fig:ranking-efficiency}(c--f)).
GPT-5.6 Sol and GPT-5.6 Terra achieve similar overall scores and runtimes (227 versus 239\,s/task), but GPT-5.6 Terra uses fewer tokens (298.7K versus 349.2K), the fewest turns of any model (18.3/task), and roughly one-third the estimated cost (\$0.68 versus \$1.97/task).
GPT-5.6 Luna offers the best score--cost trade-off, ranking third overall (68.87) at just \$0.07/task.
Despite using a similar number of tokens as GPT-5.6 Terra (285.5K versus 298.7K), lower token pricing cuts its cost by roughly ten times, though its runtime is longer (435\,s/task, 1.8$\times$ slower).
Cost and runtime thus favor different top models.

\subsection{RQ4: Failure Analysis}
Across the 340 \benchetwo runs, failures happen at three successive stages: validation, output completeness, and scientific accuracy.
First, 46 runs (13.5\%) fail an execution or physical validation check.
Second, among the 294 runs passing both checks, 72 (24.5\%) still fall short of full-completion requirements, leaving required outputs incomplete.
Third, clean execution still allows scientifically inaccurate results, as seen in indexing runs that pass validation yet recover inaccurate unit cells or reflection assignments.

These failure modes point to two priorities for agent design: adding a pre-submission check to verify all required files and supporting evidence before finishing, and checking candidate indexing solutions by comparing the predicted peak positions of competing unit cells against the measured pattern to flag ambiguous fits.

\subsection{RQ5: Component Ablation}
\begin{table}[!t]
\caption{Component Ablations on \benchqa. Relative gains use the score without the component as the baseline.}
\label{tab:component-ablation}
\centering
\small
\setlength{\tabcolsep}{2.4pt}
\begin{tabular}{p{\dimexpr\columnwidth-133pt\relax}
                S[table-format=2.2,table-column-width=32pt]
                S[table-format=2.2,table-column-width=23pt]
                S[table-format=+2.1,table-column-width=27pt,retain-explicit-plus=true]
                S[table-format=+2.1,table-column-width=27pt,retain-explicit-plus=true]}
\toprule
\rowcolor{blue!6}
\textbf{Component} & {\textbf{W/O}} & {\textbf{With}} & {\textbf{$\Delta$ (pt)}} & {\textbf{$\Delta$ (\%)}} \\
\midrule
Result acceptance checks & 72.50 & 90.00 & +17.5 & +24.1 \\
Final scientific review  & 62.71 & 79.80 & +17.1 & +27.2 \\
Domain procedures        & 66.66 & 69.14 & +2.5  & +3.7 \\
Pattern QC               & 74.07 & 75.65 & +1.6  & +2.1 \\
Residual reasoning       & 88.35 & 89.62 & +1.3  & +1.4 \\
Staged refinement        & 40.55 & 41.68 & +1.1  & +2.8 \\
\bottomrule
\end{tabular}
\end{table}

Removing any of the six evaluated AutoXRD components lowers GPT-5.6 Luna's mean score on the matched \benchqa subset (Table~\ref{tab:component-ablation}), confirming that every module contributes positively.
The gains fall into two clear groups.
Explicit verification and review checks bring the largest improvements: result-acceptance checks yield the largest absolute gain (+17.5 points, or +24.1\%, from 72.50 to 90.00), and final scientific review yields the largest relative gain (+27.2\%, or +17.1 points, from 62.71 to 79.80).
By contrast, the other four workflow modules provide smaller, consistent gains of 1.1--2.5 points, namely domain procedures (+2.5), pattern QC (+1.6), residual reasoning (+1.3), and staged refinement (+1.1).
Consistent with RQ1 and RQ4, these results show that explicit verification checks provide the strongest protection against scientific errors.

\section{Conclusion}
We introduced \method, an LLM agent framework for powder diffraction combining domain planning, crystallographic tools, and rule-based validation, along with \bench, a two-track benchmark of 100 diagnostic (\benchqa) and 34 end-to-end tasks (\benchetwo).
Across 1,340 runs of ten LLMs, diagnostic reasoning only moderately predicts end-to-end execution: models review completed fits far better than they manage coupled parameters or quantitative tasks.
On \benchetwo, indexing and peak analysis run cleanly yet remain inaccurate, whereas Rietveld refinement is limited by validation failures and incomplete outputs; ablations confirm that explicit verification checks drive the largest gains.

Reliable diffraction agents therefore cannot rely on LLM reasoning or scalar metrics such as $\rwp$ alone, and instead need to pair planning with physical constraints, completeness checks, and verifiable run records.
Future work should strengthen agents' ability in indexing candidate search, coupled-parameter control, and planning efficiency, while reporting validation, completeness, and cost alongside accuracy.

\section*{Ethical Considerations}
\bench uses only public inorganic diffraction data and open crystallographic software in compliance with their respective licenses. Because unverified structural conclusions can affect downstream materials experiments, \method is designed as an auditable decision-support tool with explicit validation checks rather than a replacement for expert scientific review.

\bibliographystyle{ACM-Reference-Format}
\bibliography{ref}

\clearpage
\onecolumn
\appendix
\raggedbottom

\section{Benchmark Resources}
\label{app:resources}
We describe the diffraction inputs and the XRD skills and tools available to the agent.

\subsection{Data sources}
\benchqa uses 70 generated cases and 30 cases based on experimental patterns. Each of the 34 \benchetwo workflows uses a distinct measured pattern. The ``34 real'' entry under Mtls.\ in Table~\ref{tab:related-comparison} counts only these 30 X-ray and four neutron patterns. The 30 experimental QA cases contribute to the 134-task total but not to the materials count. The agent receives the task and its inputs; the evaluator uses the scoring targets and rubrics.

\begin{table}[H]
\caption{Sources of the 100 QA tasks and 34 end-to-end workflows.}
\label{tab:source-provenance}\centering\small
\rowcolors{2}{tablerow}{white}
\begin{tabularx}{\textwidth}{L{3.2cm}C{1.05cm}C{0.9cm}Y}
\toprule\rowcolor{tablehead}
\textbf{Source} & \textbf{Track} & \textbf{Tasks} & \textbf{Evidence} \\
Generated cases & QA & 70 & Synthetic patterns, residual summaries, and refinement histories \\
IUCr CPD round robin & QA & 20 & Measured patterns for interpretation and quantitative phase analysis \\
Dara & QA & 10 & Measured patterns and candidate phase lists for phase identification \\
opXRD & E2E & 30 & Laboratory Cu K$\alpha$ powder patterns from two collections \\
FullProf examples & E2E & 4 & Two time-of-flight and two constant-wavelength neutron patterns \\
\bottomrule\end{tabularx}
\end{table}

\Needspace{205pt}
\subsection{XRD procedure skills}
The skills guide input checks, staged refinement, and result validation.

\begin{table}[H]
\caption{The ten XRD procedure skills in \method.}
\label{tab:ten-skills}\centering\small
\rowcolors{2}{tablerow}{white}
\begin{tabularx}{\textwidth}{L{5.2cm}Y}
\toprule\rowcolor{tablehead}\textbf{Skill} & \textbf{Scientific role} \\
\path{fullprof-le-bail} & Guides Le Bail whole-pattern decomposition. \\
\path{fullprof-pcr-compiler} & Compiles structured FullProf inputs and checks parameter selectors. \\
\path{fullprof-staged-refinement} & Sequences parameter release and stage checks. \\
\path{gsasii-executable-workflow} & Guides GSAS-II project setup, fitting, and export. \\
\path{xrd-pattern-qc} & Reviews range, sampling, and signal quality. \\
\path{xrd-physical-audit} & Checks refined models for physical plausibility. \\
\path{xrd-residual-diagnosis} & Links residual signatures to mechanisms and actions. \\
\path{xrd-residual-features} & Computes summary and local residual features. \\
\path{xrd-structure-audit} & Reviews symmetry, occupancies, and distances. \\
\path{xrd-trajectory-gate} & Checks candidate transitions against predicted changes. \\
\bottomrule\end{tabularx}
\end{table}

\subsection{Agent tools}

\begin{table}[H]
\caption{Tools in \method.}
\label{tab:agent-tools}\centering\small
\rowcolors{2}{tablerow}{white}
\begin{tabularx}{\textwidth}{L{3.5cm}Y}
\toprule\rowcolor{tablehead}\textbf{Tool} & \textbf{Purpose} \\
\rowcolor{white}\multicolumn{2}{l}{\textit{Native tools}} \\
\path{Read} & Reads files in the workspace. \\
\path{Glob} & Finds files matching a path pattern. \\
\path{Grep} & Searches file contents for a pattern. \\
\path{Edit} & Makes targeted edits to existing files. \\
\path{Write} & Creates or replaces files. \\
\path{Bash} & Runs commands and analysis scripts in the workspace. \\
\midrule
\rowcolor{white}\multicolumn{2}{l}{\textit{XRD utilities}} \\
\path{xrd.pattern} & Reads powder patterns and checks scan range, sampling, and intensity. \\
\path{xrd.structure} & Reads CIFs and checks unit cells, site occupancies, and interatomic distances. \\
\path{xrd.fullprof} & Runs FullProf and extracts fit statistics and convergence information. \\
\path{xrd.le_bail} & Validates the input template, runs Le Bail profile matching, and checks for a reflection file. \\
\path{xrd.pcr} & Inspects FullProf input files and writes a new PCR file with the selected parameters released. \\
\path{xrd.residual} & Extracts features such as peak mismatches and baseline trends from observed-minus-calculated profiles. \\
\path{xrd.trajectory} & Checks proposed actions, evaluates before-and-after fit states, and stores refinement histories. \\
\path{pattern_qc.py} & Reports sampling, noise, and candidate peak positions from two-column patterns. \\
\path{validate_result.py} & Checks fit statistics, site occupancies, isotropic displacement parameters, and phase fractions in parsed results. \\
\bottomrule\end{tabularx}
\end{table}

\Needspace{220pt}
\section{Task Coverage}
\label{app:coverage}
The tables organize tasks by QA capability and end-to-end reporting domain.

\begin{table}[H]
\caption{QA capabilities and task counts (Figure~\ref{fig:combined-radar}).}
\label{tab:qa-capabilities}\centering\small
\rowcolors{2}{tablerow}{white}
\begin{tabularx}{\textwidth}{L{1.35cm}L{4.0cm}C{0.9cm}Y}
\toprule\rowcolor{tablehead}\textbf{Tier} & \textbf{Capability} & \textbf{Tasks} & \textbf{Decision or result} \\
Easy & Refinement-action selection & 10 & Assess whether a proposed action is admissible. \\
Easy & Result acceptance & 10 & Decide whether a candidate passes validation. \\
Easy & Residual diagnosis & 10 & Identify the residual mechanism and next action. \\
Medium & Experimental interpretation & 10 & Interpret a measured pattern and propose verification. \\
Medium & Residual interpretation & 20 & Explain the residual evidence and recommend an action. \\
Medium & Refinement-history assessment & 10 & Review a trajectory and decide whether to stop. \\
Hard & Parameter recovery & 10 & Estimate correction parameters and plan staged refinement. \\
Hard & Phase quantification & 10 & Estimate supported phases and their weight fractions. \\
Hard & Phase identification & 10 & Identify phases and quantify confidence. \\
\bottomrule\end{tabularx}
\end{table}

\begin{table}[H]
\caption{Eleven end-to-end workflow families mapped to nine reporting domains (Figure~\ref{fig:combined-radar}).}
\label{tab:e2e-domains}\centering\small
\rowcolors{2}{tablerow}{white}
\begin{tabularx}{\textwidth}{L{4.1cm}YC{0.9cm}}
\toprule\rowcolor{tablehead}\textbf{Reporting domain} & \textbf{Workflow families} & \textbf{Tasks} \\
Search \& phase ID & Phase identification (4), database retrieval (2) & 6 \\
Phase quantification & Phase quantification & 5 \\
Structure selection & Structure determination & 3 \\
Rietveld refinement & Rietveld refinement & 8 \\
Indexing & Indexing & 2 \\
Peak analysis & Peak analysis & 3 \\
Microstructure \& disorder & Crystallite analysis (2), defect/disorder (2) & 4 \\
Pattern prediction & Pattern prediction & 2 \\
Property inference & Material property prediction & 1 \\
\bottomrule\end{tabularx}
\end{table}

\section{Representative Benchmark Cases}
\label{sec:benchmark-case-examples}
Six examples illustrate how the supplied evidence, required outputs, and scoring differ between the two tracks.

\subsection{\benchqa}
The three QA cases ask for a refinement decision, a diagnosis from residual evidence, and a quantitative phase estimate.

\begin{xrdcasebox}{Checking an Fe occupancy action}{Easy \enspace\texttt{easy-action-004}}
\casefield{Input}{A structure-stage action proposes refining the Fe site occupancy without explicit bounds. It cites a residual score of 0.30 and predicts a decrease of at least 0.05.}
\casefield{Task}{Select all valid statements about whether the proposed action is admissible, including whether its refinement stage, bounds, and isolation from other mechanisms are appropriate. Submit the selected option IDs.}
\casefield{Evaluation}{The selected option set is scored by exact match against the reference from the deterministic action validator.}
\end{xrdcasebox}

\begin{xrdcasebox}{Diagnosing a CeO$_2$ residual}{Medium \enspace\texttt{medium-residual-001}}
\casefield{Input}{A synthetic CeO$_2$ profile provides 2,001 samples of $2\theta$, observed intensity, and calculated intensity.}
\casefield{Task}{Identify the dominant residual mechanism from the observed-minus-calculated profile. Cite a diagnostic metric or feature, recommend the next isolated refinement action, and explain why one plausible alternative is inappropriate.}
\casefield{Evaluation}{The scientific judge assesses the diagnosis, its supporting evidence, the proposed action, and the reasoning used to rule out the alternative.}
\end{xrdcasebox}

\begin{xrdcasebox}{Quantifying an experimental mixture}{Hard \enspace\texttt{hard-qpa-001}}
\casefield{Input}{A measured Cu K$\alpha$ pattern for sample 1a of the IUCr round robin and a list of ten candidate phases.}
\casefield{Task}{Identify the phases supported by the pattern, estimate weight fractions that sum to one, and report per-phase uncertainties. Explain the evidence and any dominant artifact in a short report.}
\casefield{Evaluation}{The evaluator scores phase-set F1, mean absolute and root-mean-square errors against the weighed fractions, and the artifact label. The scientific judge assesses interpretation and uncertainty.}
\end{xrdcasebox}

\Needspace{185pt}
\subsection{\benchetwo}
These workflows illustrate peak fitting, powder indexing, and staged neutron refinement.
Each requires an executed analysis, \texttt{final\_report.md}, and the execution record \path{artifacts/execution.json}.

\begin{xrdcasebox}[e2ecase]{Fitting peaks in RuO$_2$}{Medium \enspace\texttt{e2e-v3-009}}
\casefield{Input}{A measured Cu K$\alpha$ profile with 6,000 samples (\texttt{pattern.xye}) and a starting instrument model (\texttt{instrument.prm}).}
\casefield{Task}{Detect and fit peaks, separating overlapping components where the data support it. Save positions, integrated intensities, full widths at half maximum, position uncertainties, and unresolved-component flags in \texttt{peaks.csv}. Save the observed, fitted, and difference curves in \texttt{fit\_curve.csv}.}
\casefield{Evaluation}{The numerical score checks peak detection, position and width errors, intensity ordering, and the quality of the reconstructed profile.}
\end{xrdcasebox}

\begin{xrdcasebox}[e2ecase]{Indexing a WO$_3$ powder pattern}{Hard \enspace\texttt{e2e-v3-011}}
\casefield{Input}{A measured Cu K$\alpha$ pattern spanning $10^\circ$--$80^\circ$ in $2\theta$ (\texttt{pattern.xye}) and the instrument parameters (\texttt{instrument.prm}). No candidate structure is supplied.}
\casefield{Task}{Recover and refine a unit cell from the measured peak positions, assign $(hkl)$ indices, and check the indexing against all significant peaks. Save the cell, figure of merit, and alternatives in \texttt{indexing.json}, and the assignments and unindexed peaks in \texttt{indexed\_peaks.csv}.}
\casefield{Evaluation}{The numerical score evaluates agreement of cell invariants, peak-match F1, peak-position errors, and consistency of the reflection assignments.}
\end{xrdcasebox}

\begin{xrdcasebox}[e2ecase]{Two-phase neutron Rietveld refinement}{Hard \enspace\texttt{e2e-v3-031}}
\casefield{Input}{An OSIRIS time-of-flight neutron profile (\texttt{pattern.dat}) and an executable FullProf starting model (\texttt{starting\_state.pcr}) containing Na$_2$Ca$_3$Al$_2$F$_{14}$ and CaF$_2$.}
\casefield{Task}{Run and revise the FullProf model in stages. Submit the final PCR and matching backend output, refined cell and fit statistics, observed and calculated profiles with their difference, the accepted and rejected steps, and physical checks.}
\casefield{Evaluation}{The numerical criteria include recomputed $R_{\mathrm{wp}}$, profile errors, cell invariants, residual structure, the quality of the refinement history, and whether refinement updated the starting state.}
\end{xrdcasebox}

\section{Agent and Evaluator Prompts}
\label{app:prompts}
The boxes below present condensed instructions for the agents and judges.

\begin{promptbox}{QA agent instructions}
\small\ttfamily
Inspect the supplied evidence with the tools and XRD skills in the case workspace.\\[2pt]
Easy: return a JSON object in the required schema.\\
Medium/Hard: give a plain-text scientific report with the conclusion, quantitative results, supporting evidence, uncertainty, and recommended action.
\end{promptbox}

\begin{promptbox}[colback=autoorangelight,colframe=autoorange,colbacktitle=autoorange]{End-to-end agent instructions}
\small\ttfamily
Inspect the input files. Use Read, Glob, Grep, Edit, Write, and Bash to prepare and run a crystallographic or numerical analysis in the case workspace. Inspect the outputs and revise the workflow as needed.\\[2pt]
Submit the executed code, declared artifacts, and \texttt{final\_report.md}. Document the commands, results, residual and physical checks, and uncertainties.
\end{promptbox}

\begin{promptbox}[colback=autopurplelight,colframe=autopurple,colbacktitle=autopurple]{QA scientific judge instructions}
\small\ttfamily
Score scientific correctness, evidence grounding, action quality, and completeness from 0 to 4. Credit scientifically equivalent diagnoses and actions. Mark central failures as fatal. For Hard tasks, extract the reported numerical values for deterministic scoring.
\end{promptbox}

\Needspace{105pt}
\subsection{End-to-end scientific judging}
\label{sec:e2e-judge}
GPT-5.6 Sol evaluates runs that pass the execution and physical checks. When available, scientific references are supplied to the judge and withheld from the agent.

\begin{promptbox}[colback=autopurplelight,colframe=autopurple,colbacktitle=autopurple]{End-to-end scientific judge instructions}
\small\ttfamily
Assess the reported analysis against the task requirements, generated files, and numerical checks. Use reference results as guidance and credit equivalent answers supported by the run. When references are incomplete, assess the available evidence and distinguish missing reference values from missing required results. Rate scientific correctness, evidence grounding, workflow validity, completeness, and uncertainty calibration from 0 to 4. Flag failures that invalidate the main conclusion or workflow and return the requested JSON object.
\end{promptbox}

For task $j$, the evaluator computes $\phi_{J,j}=\frac{1}{20}\sum_{d=1}^{5}s_{j,d}$ from the five ratings $s_{j,d}\in[0,4]$, giving them equal weight. A fatal-error flag caps $\phi_{J,j}$ at $0.25$. For runs with valid execution and physical results, $\phi_{J,j}$ contributes 30\% of the task score, with 15\% from output file validity and 55\% from numerical metrics. A run that fails either check receives zero and is not sent to the judge.

\begin{xrdcasebox}[e2ecase]{Example judge request and response}{\texttt{e2e-v3-031}}
\casefield{Task}{Two-phase time-of-flight neutron Rietveld refinement with FullProf (Appendix~\ref{sec:benchmark-case-examples}). This run was submitted by GPT-5.6 Terra.}
\casefield{Required files}{\path{final_report.md} and seven files in \path{artifacts/}: \path{execution.json}, \path{final_project.pcr}, \path{final_backend.out}, \path{refinement_metrics.json}, \path{difference_curve.csv}, \path{trajectory.json}, and \path{physical_checks.json}.}
\casefield{Soft ground truth (reference solution)}{The reference phases are Na$_2$Ca$_3$Al$_2$F$_{14}$ and CaF$_2$, with respective cubic cell edges of 10.2515 and 5.4646\,\AA\ and phase fractions of 0.9938 and 0.0062. The rubric credits equivalent parameterizations supported by an executed refinement.}
\casefield{Numerical checks}{Execution and physical checks both pass. The evaluator recomputes $R_{\mathrm{wp}}=33.8\%$, a normalized profile RMSE of 0.01969, and a cell-invariant error of $1.95\times10^{-6}$.}
\casefield{Judge response}{GPT-5.6 Sol assigns 3 for scientific correctness and 4 for each remaining dimension, yielding $\phi_{J,j}=19/20=0.95$ with no fatal error. The critique credits three documented FullProf runs and recognizes that fixing the Ca displacement parameter in CaF$_2$ is a modeling assumption. It also notes that structured residuals and an ill-conditioned fit limit the precision of the minor-phase estimates.}
\end{xrdcasebox}

\section{Experimental Protocol}
\label{app:protocol}
The protocol specifies the interaction budgets, scientific judges, and resampling procedure used for the reported scores.

The reported QA runs allow 20 tool calls per task and 8,192 output tokens per turn, with GPT-5.6 Luna as the scientific judge. The corresponding end-to-end limits are 50 agent steps and 32,768 output tokens; GPT-5.6 Sol judges the reports. To limit potential judge bias, deterministic checks dominate scoring in both tracks: on \benchqa, 30 tasks use exact matching and Hard tasks score numerical outputs deterministically, with GPT-5.6 Luna ranking third on \benchqa; on \benchetwo, the scientific judge contributes 30\% of the score only after execution and physical checks pass, and GPT-5.6 Sol ranks second behind GPT-5.6 Terra on \benchetwo.

The overall score averages the two track scores. Its 95\% confidence intervals use 20,000 task-level bootstrap replicates within tracks, shared across models for pairwise comparisons.

\subsection{Run accounting}
We report totals for evaluated runs, model turns, and tokens separately for each track.

\begin{table}[H]
\caption{Evaluated runs and resource totals across the ten models.}
\label{tab:run-totals}\centering\small
\rowcolors{2}{tablerow}{white}
\begin{tabularx}{\textwidth}{Y*{3}{Z}}
\toprule\rowcolor{tablehead}\textbf{Track} & \textbf{Runs} & \textbf{Model turns} & \textbf{Tokens (millions)} \\
\benchqa & 1,000 & 9,794 & 198.2 \\
\benchetwo & 340 & 12,341 & 427.4 \\
\midrule
Total & 1,340 & 22,135 & 625.6 \\
\bottomrule\end{tabularx}
\end{table}

\Needspace{235pt}
\section{Detailed Results}
\label{app:results}
We report scores by model and task category, followed by component ablations.

\subsection{Model scores}
On point estimates, GPT-5.6 Sol leads overall and GPT-5.6 Terra leads the end-to-end track.

\begin{table}[H]
\caption{Model scores with 95\% bootstrap intervals. The last column counts end-to-end runs passing both execution and physical checks.}
\label{tab:unified-results}\centering\small\setlength{\tabcolsep}{4pt}
\rowcolors{2}{tablerow}{white}
\begin{tabularx}{\textwidth}{L{3.2cm}YYYY}
\toprule\rowcolor{tablehead}\textbf{Model} & \textbf{QA [95\% CI]} & \textbf{E2E [95\% CI]} & \textbf{Overall [95\% CI]} & \textbf{E2E validated} \\
GPT-5.6 Sol & \textbf{82.2 [76.1, 87.7]} & 80.0 [73.8, 85.4] & \textbf{81.1 [76.8, 85.0]} & \textbf{34/34} \\
GPT-5.6 Terra & 77.8 [71.2, 84.1] & \textbf{81.0 [76.5, 85.2]} & 79.4 [75.4, 83.3] & \textbf{34/34} \\
GPT-5.6 Luna & 72.2 [65.1, 79.0] & 65.5 [56.0, 74.3] & 68.9 [63.0, 74.6] & 31/34 \\
DeepSeek-V4-Pro & 57.2 [48.6, 65.8] & 73.7 [67.3, 79.9] & 65.5 [60.0, 70.8] & \textbf{34/34} \\
MiMo-V2.5-Pro & 69.0 [62.2, 75.5] & 47.2 [37.5, 56.6] & 58.1 [52.2, 63.9] & 33/34 \\
MiMo-V2.5 & 64.4 [57.0, 71.6] & 38.8 [27.9, 49.7] & 51.6 [44.9, 58.2] & 25/34 \\
DeepSeek-V4-Flash & 63.4 [55.4, 71.0] & 35.8 [24.9, 47.3] & 49.6 [42.8, 56.5] & 24/34 \\
Gemini 3.1 Pro Low & 36.5 [27.8, 45.5] & 57.1 [49.6, 64.1] & 46.8 [41.1, 52.4] & 33/34 \\
MiniMax-M2.7 & 50.6 [42.8, 58.3] & 30.2 [21.3, 39.6] & 40.4 [34.5, 46.5] & 25/34 \\
Claude Haiku 4.5 & 45.4 [37.4, 53.4] & 27.4 [19.1, 35.8] & 36.4 [30.7, 42.2] & 21/34 \\
\bottomrule\end{tabularx}
\end{table}

\subsection{Capability profiles}
These profiles break down model scores by diagnostic capability and workflow domain.

\begin{table}[H]
\caption{QA scores by capability (Figure~\ref{fig:combined-radar}). RD: residual diagnosis; EI: experimental interpretation; RI: residual interpretation; RH: refinement-history assessment; PR: parameter recovery; QPA: phase quantification; PID: phase identification.}
\label{tab:qa-matrix}\centering\scriptsize\setlength{\tabcolsep}{3.1pt}
\rowcolors{2}{tablerow}{white}
\begin{tabularx}{\textwidth}{L{2.8cm}*{9}{Z}}
\toprule\rowcolor{tablehead}\textbf{Model} & \textbf{Action} & \textbf{Accept} & \textbf{RD} & \textbf{EI} & \textbf{RI} & \textbf{RH} & \textbf{PR} & \textbf{QPA} & \textbf{PID} \\
GPT-5.6 Sol & 60.0 & 80.0 & 100.0 & 83.1 & 92.2 & 92.5 & 69.6 & 68.2 & 84.1 \\
GPT-5.6 Terra & 60.0 & 90.0 & 90.0 & 73.8 & 91.9 & 92.5 & 59.3 & 61.4 & 67.6 \\
GPT-5.6 Luna & 50.0 & 80.0 & 80.0 & 61.2 & 83.1 & 92.5 & 69.3 & 44.5 & 78.3 \\
DeepSeek-V4-Pro & 50.0 & 80.0 & 90.0 & 60.6 & 45.6 & 91.2 & 54.9 & 4.8 & 49.5 \\
MiMo-V2.5-Pro & 50.0 & 80.0 & 80.0 & 56.2 & 73.1 & 85.0 & 65.1 & 50.1 & 77.7 \\
MiMo-V2.5 & 30.0 & 80.0 & 70.0 & 55.0 & 66.6 & 86.2 & 55.7 & 58.0 & 76.4 \\
DeepSeek-V4-Flash & 60.0 & 70.0 & 70.0 & 70.6 & 78.4 & 88.8 & 58.0 & 7.2 & 52.3 \\
Gemini 3.1 Pro Low & 50.0 & 90.0 & 10.0 & 21.2 & 28.1 & 88.8 & 28.7 & 0.0 & 20.3 \\
MiniMax-M2.7 & 10.0 & 80.0 & 60.0 & 45.0 & 57.5 & 74.4 & 32.3 & 29.6 & 59.7 \\
Claude Haiku 4.5 & 20.0 & 80.0 & 40.0 & 34.4 & 45.0 & 88.1 & 31.6 & 6.6 & 63.8 \\
\midrule\textbf{Mean} & \textbf{44.0} & \textbf{81.0} & \textbf{69.0} & \textbf{56.1} & \textbf{66.2} & \textbf{88.0} & \textbf{52.4} & \textbf{33.0} & \textbf{63.0} \\
\bottomrule\end{tabularx}
\end{table}

\begin{table}[H]
\caption{End-to-end scores by reporting domain (Figure~\ref{fig:combined-radar}). The workflow families for each domain are listed in Table~\ref{tab:e2e-domains}.}
\label{tab:e2e-matrix}\centering\scriptsize\setlength{\tabcolsep}{2.5pt}
\rowcolors{2}{tablerow}{white}
\begin{tabularx}{\textwidth}{L{2.8cm}*{9}{Z}}
\toprule\rowcolor{tablehead}\textbf{Model} & \textbf{Search} & \textbf{QPA} & \textbf{Struct.} & \textbf{Rietveld} & \textbf{Index} & \textbf{Peaks} & \textbf{Micro.} & \textbf{Pattern} & \textbf{Property} \\
GPT-5.6 Sol & 76.0 & 91.6 & 90.2 & 80.3 & 56.5 & 61.4 & 78.8 & 91.4 & 97.9 \\
GPT-5.6 Terra & 78.0 & 92.5 & 85.7 & 78.1 & 69.3 & 64.1 & 84.0 & 89.9 & 94.6 \\
GPT-5.6 Luna & 74.1 & 51.9 & 82.0 & 61.3 & 37.2 & 67.7 & 58.6 & 91.3 & 93.2 \\
DeepSeek-V4-Pro & 72.4 & 87.7 & 78.6 & 72.0 & 53.4 & 54.9 & 69.5 & 87.9 & 94.9 \\
MiMo-V2.5-Pro & 40.1 & 41.0 & 51.5 & 49.8 & 36.5 & 62.0 & 27.1 & 77.0 & 85.2 \\
MiMo-V2.5 & 62.1 & 36.2 & 25.8 & 19.5 & 37.1 & 65.2 & 38.6 & 40.5 & 28.1 \\
DeepSeek-V4-Flash & 53.7 & 41.8 & 41.9 & 18.1 & 35.4 & 50.1 & 28.2 & 41.0 & 0.0 \\
Gemini 3.1 Pro Low & 62.3 & 45.4 & 70.7 & 59.3 & 28.2 & 56.2 & 51.0 & 67.4 & 90.5 \\
MiniMax-M2.7 & 32.2 & 47.6 & 18.3 & 23.1 & 23.7 & 46.7 & 28.0 & 9.4 & 35.5 \\
Claude Haiku 4.5 & 29.0 & 43.6 & 30.3 & 24.3 & 0.0 & 47.5 & 15.3 & 11.2 & 28.1 \\
\bottomrule\end{tabularx}
\end{table}

\Needspace{205pt}
\subsection{End-to-end workflow results}
The results separate validation, deliverable completion, and scientific accuracy.

\begin{table}[H]
\caption{Results by workflow family across ten models (Figure~\ref{fig:e2e-bottlenecks}). $n$ is the number of workflows. Validation and completion are percentages. Component means include zero credit for runs that fail validation.}
\label{tab:e2e-bottlenecks}\centering\small\setlength{\tabcolsep}{2pt}
\rowcolors{2}{tablerow}{white}
\begin{tabularx}{\textwidth}{L{3.6cm}r*{6}{Z}}
\toprule\rowcolor{tablehead}\textbf{Family} & \textbf{$n$} & \textbf{Final} & \textbf{Validation} & \textbf{Complete} & \textbf{Files} & \textbf{Metrics} & \textbf{Judge} \\
Property prediction & 1 & 64.79 & 90.00 & 70.00 & 85.0 & 64.4 & 55.5 \\
Database retrieval & 2 & 62.03 & 90.00 & 70.00 & 78.8 & 59.9 & 57.5 \\
Pattern prediction & 2 & 60.70 & 85.00 & 70.00 & 81.2 & 58.2 & 55.0 \\
Phase quantification & 5 & 57.92 & 86.00 & 56.00 & 69.6 & 58.1 & 51.7 \\
Peak analysis & 3 & 57.60 & 96.67 & 90.00 & 94.2 & 54.0 & 45.8 \\
Structure determination & 3 & 57.52 & 96.67 & 70.00 & 83.3 & 52.0 & 54.7 \\
Phase identification & 4 & 55.97 & 97.50 & 75.00 & 86.2 & 48.7 & 54.2 \\
Rietveld refinement & 8 & 48.58 & 71.25 & 50.00 & 66.2 & 48.6 & 39.8 \\
Crystallite analysis & 2 & 48.30 & 85.00 & 65.00 & 77.5 & 40.8 & 47.5 \\
Defect/disorder & 2 & 47.52 & 90.00 & 65.00 & 71.2 & 40.5 & 48.5 \\
Indexing & 2 & 37.72 & 90.00 & 80.00 & 86.2 & 26.9 & 33.2 \\
\bottomrule\end{tabularx}
\end{table}

\Needspace{180pt}
\subsection{Component ablations}
For each component, we compare the full system and the ablated system on the same QA cases.

\begin{table}[H]
\caption{GPT-5.6 Luna ablations on paired QA cases (Table~\ref{tab:component-ablation}). Gains are differences between the unrounded full and removal scores.}
\label{tab:app-component-ablation}\centering\small
\rowcolors{2}{tablerow}{white}
\begin{tabularx}{\textwidth}{L{5.5cm}*{4}{Z}}
\toprule\rowcolor{tablehead}\textbf{Component} & \textbf{Cases} & \textbf{Full} & \textbf{Removed} & \textbf{Gain} \\
Result acceptance checks & 10 & 90.00 & 72.50 & +17.5 \\
Final scientific review & 10 & 79.80 & 62.71 & +17.1 \\
Domain procedures & 10 & 69.14 & 66.66 & +2.5 \\
Pattern QC & 20 & 75.65 & 74.07 & +1.6 \\
Residual reasoning & 10 & 89.62 & 88.35 & +1.3 \\
Staged refinement & 10 & 41.68 & 40.55 & +1.1 \\
\bottomrule\end{tabularx}
\end{table}

\end{document}